\documentclass[conference]{IEEEtran}
\IEEEoverridecommandlockouts
\usepackage{authblk}
\usepackage{cite}
\usepackage{amsmath,amssymb,amsfonts}
\usepackage{adjustbox}
\usepackage{algorithmic}
\usepackage{caption}
\usepackage{multicol}
\usepackage{import}
\usepackage{balance}
\usepackage{algorithmic}
\usepackage{graphicx}
\usepackage[colorlinks,
    linkcolor=blue,
    citecolor=blue,
    pdftex,
    pdfauthor={assert},
    pdftitle={rick},
    pdfsubject={unicorns and apis},
    pdfkeywords={rickroll, execution},
    pdfproducer={mel brooks},
    pdfcreator={gru}]{hyperref}
\usepackage[dvipsnames, table]{xcolor}
\usepackage{tabularx}
\usepackage{textcomp}
\usepackage{xcolor}
\usepackage{xspace}
\usepackage{booktabs}
\definecolor{light-gray}{gray}{0.95}
\definecolor{pgreen}{RGB}{5,205,107}
\definecolor{pblue}{RGB}{2,154,223}
\usepackage{multirow}
\usepackage{listings}
\lstdefinestyle{java}{
    basicstyle=\ttfamily\scriptsize, 
    stringstyle=\bfseries\color{teal},
    commentstyle=\bfseries\color{purple}
}

\newcommand{\lstbg}[3][0pt]{{\fboxsep#1\colorbox{#2}{\strut #3}}}
\definecolor{codegreen}{rgb}{0,0.6,0}
\lstdefinelanguage{diff}{
    basicstyle=\ttfamily\scriptsize,
	morecomment=[f][\color{red}]{---}, 
	morecomment=[f][\color{codegreen}]{+++},
	morecomment=[f][\lstbg{red!20}]{-\ },
	morecomment=[f][\lstbg{green!20}]{\ +\ },
	morecomment=[f][\color{blue}]{@@},}
	
\makeatletter
\lst@AddToHook{OnEmptyLine}{\addtocounter{lstnumber}{-1}}
\makeatother

\newcommand{\squeezeup}{\vspace{-2.2mm}}

\def\BibTeX{{\rm B\kern-.05em{\sc i\kern-.025em b}\kern-.08em
    T\kern-.1667em\lower.7ex\hbox{E}\kern-.125emX}}

\newcommand{\rick}{\textsc{rick}\xspace}

\newcommand{\graphhopper}{GraphHopper\xspace}

\title{\rick: Generating Mocks from Production Data}

\author{Deepika Tiwari, Martin Monperrus, Benoit Baudry\\
KTH Royal Institute of Technology, Sweden\\
\{deepikat, monperrus, baudry\}@kth.se}

\begin{document}

\maketitle
\begin{abstract}
Test doubles, such as mocks and stubs, are nifty fixtures in unit tests.
They allow developers to test individual components in isolation from others that lie within or outside of the system.
However, implementing test doubles within tests is not straightforward.
With this demonstration, we introduce \rick, a tool that observes executing applications in order to automatically generate tests with realistic mocks and stubs. \rick monitors the invocation of target methods and their interactions with external components. Based on the data collected from these observations, \rick produces unit tests with mocks, stubs, and mock-based oracles.
We highlight the capabilities of \rick, and how it can be used with real-world Java applications, to generate tests with mocks.
\end{abstract}


\squeezeup

\section{Introduction}
A software application is a collection of multiple systems coupled together to achieve the goals of the whole. This facilitates incremental and modular development, as well as the reuse of external libraries. However, in order for each component to be tested on its own, it must be decoupled from the others. A solution to this problem is the use of test doubles within unit tests. External components are replaced with skeletal implementations called \emph{mocks} \cite{mackinnon2000endo}. Since a mock substitutes an actual component, all interactions with the component are bypassed. Instead, the behavior of the mock is declared through \emph{stubs}, so that when a method is triggered with a certain input on a mock, a canned response is returned.
Mocks and stubs allow for more focus on verifying the correctness of the component under test, fewer side effects, faster test executions, and quicker fault localization \cite{thomas2002mock}. However, using them is not trivial. When writing a unit test, developers must manually decide which components may be mocked, and also how they are stubbed~\cite{barr2014oracle}.

Mocks generated automatically through search based algorithms \cite{arcuri2015generating} or symbolic execution \cite{islam2010dsc+} may not correspond to real production behavior. Our key insight is that using data collected in production to generate tests with mocks ensures that behavior triggered by actual users is tested, and can even complement the developer-written test suite \cite{9526340}.
We propose \rick, a tool that observes executing applications and collects data, with the goal of automatically generating tests that use mocks, stubbing their behavior in accordance with production invocations. Furthermore, the tests generated by \rick contain \emph{mock-based oracles}, derived from data collected in production, that verify distinct aspects of the invocation of a target method and its interactions with external components.

This demonstration presents an overview of \rick. Our companion video \cite{demo} illustrates \rick in action with the open-source routing application called \graphhopper \cite{graphhopper}.

\section{\rick}
This section highlights the key phases of \rick, and presents examples of generated tests. We also summarize how \rick can be used with real-world Java applications.

\subsection{Key phases for mock generation} 

\begin{lstlisting}[style=java, label={lst:mut}, caption={An MUT, with two mockable method calls, identified as a target for test generation by \rick}, float, belowskip=-2.5em]
class ClassUnderTest {
  ExtTypeOne extField;
  ...
  int methodUnderTest(int a, ExtTypeTwo extParam) {
    ...
    int x = extField.mockableMethodOne(a);
    int y = extParam.mockableMethodTwo(b);
    ...
    return ...;
  }
}
\end{lstlisting}

\rick generates tests and mocks in three phases. First,  \rick identifies a configurable set of target methods for test generation. We call each target method a \emph{method under test} or an \emph{MUT}. An MUT has \emph{mockable method calls} on objects of external types, where these types may be defined as fields in the declaring type of the MUT, or as parameters to the MUT. For example, \autoref{lst:mut} presents an MUT called \texttt{methodUnderTest} defined in the class \texttt{ClassUnderTest}. There are two mockable method calls within the MUT, one on the field \texttt{extField} (line $6$), and the other on the parameter \texttt{extParam} (line $7$).
The second phase of \rick occurs when the application is in production. \rick collects data corresponding to each invocation of an MUT, as well as the mockable method calls that occur within it. For the MUT, the data collected includes the \emph{receiving object}, i.e., the object on which the MUT is invoked, the parameters passed to the invocation, as well as the object returned from it. For the invocations of the mockable methods, \rick captures the parameters and the returned objects.
Finally, in the third phase, \rick uses the collected data offline to generate tests for the MUT. The external objects are replaced with mocks within the test. The mockable method calls become \emph{mock method calls}, i.e., they are invoked on mocks instead of actual objects. Their behavior is stubbed based on the observations made in production. \rick generates three distinct kinds of mock-based oracles to verify production observations.

\subsection{Generated tests}
The tests generated by \rick mimic the observations made in production, through mocks, stubs, and mock-based oracles. 
For each invocation of an MUT, \rick produces three tests, each with an oracle that focuses on a distinct aspect of the invocation.
The \emph{Arrange} and \emph{Act} phases of the three generated tests are common. For the MUT in \autoref{lst:mut}, they are presented in \autoref{lst:test-common}. The receiving object is deserialized and recreated from its captured production snapshot (line $5$), and external objects are replaced with mocks (lines $6$ to $9$). Then,  mockable methods are stubbed based on production observations (lines $10$ to $12$). This is followed by the invocation of the MUT (line $14$). The \emph{Assert} phase in the test with an \emph{output oracle} verifies the output of the MUT, as was observed in production. We present it on line $16$ of \autoref{lst:test-oo}.
The \emph{Assert} phase in the test with a \emph{parameter oracle}, presented in \autoref{lst:test-po}, verifies that the mockable method calls occur with the same parameters within the test, as they did in production.
For a test with a \emph{call oracle}, the \emph{Assert} phase (\autoref{lst:test-co}) verifies that the mock method calls occur in the same sequence and the same number of times as was observed by \rick in production.
The three categories of mock-based oracles ensure that production conditions are recreated within the tests, with respect to production parameters and responses, as well as interactions with external components.

Capturing method invocations, persisting associated object snapshots, and expressing this information through oracles, is not trivial. We also appreciate that correctly generated tests on disk do not guarantee the faithful reproduction of production contexts. An object may not be recreated to its production state upon deserialization, and the behaviour of an MUT may be modified due to the introduction of mocks. Developing \rick and using it with complex, real-world projects has been a fruitful exercise in understanding the dynamic behaviour of software.

\begin{lstlisting}[style=java, label={lst:test-common}, caption={The arrange and act phases of a \rick test}, float, belowskip=-1em]
@Test
public void testMethodUnderTest() {
// Arrange
  // Recreate object from serialized production data
  ClassUnderTest productionObj = deserialize(new File( "receiving.xml"));
  // Inject mock field
  ExtTypeOne mockExtField = injectMockField_extField_InClassUnderTest();
  // Mock parameter
  ExtTypeTwo mockExtParam = mock(ExtTypeTwo.class);
  // Stub behavior
  when(mockExtField.mockableMethodOne(42)).thenReturn(99);
  when(mockExtParam.mockableMethodTwo(27)).thenReturn(17);
  
// Act
  int actual = productionObj.methodUnderTest(64, mockExtParam);

// Assert
  ...
\end{lstlisting}
\begin{lstlisting}[label={lst:test-oo}, caption={The assert phase of a \rick test with an output oracle}, firstnumber=16, float, belowskip=-2.5em]
  assertEquals(42, actual);
}
\end{lstlisting}

\begin{lstlisting}[label={lst:test-po}, caption={The assert phase of a \rick test with a parameter oracle}, firstnumber=16, float, belowskip=-2em]
  verify(mockExtField, atLeastOnce()).mockableMethodOne(42);
  verify(mockExtParam, atLeastOnce()).mockableMethodTwo(27);
}
\end{lstlisting}
\begin{lstlisting}[label={lst:test-co}, caption={The assert phase of a \rick test with a call oracle}, firstnumber=16, float, belowskip=-2.5em]
  InOrder orderVerifier = inOrder(mockExtField, mockExtParam);
  orderVerifier.verify(mockExtField, times(1)) .mockableMethodOne(anyInt());
  orderVerifier.verify(mockExtParam, times(1)) .mockableMethodTwo(anyInt());
}
\end{lstlisting}

\subsection{Usage}
\rick is open-source, and can be cloned from its repository on GitHub \cite{rick}. The \texttt{README} and repository \texttt{wiki} provide extensive details, with examples, on using \rick with Java applications. Our accompanying video \cite{demo} illustrates \rick in action with the open-source routing application, \graphhopper~\cite{graphhopper}.
\rick is implemented as three core modules, each invoked through simple commands. Their goals can be summarized as follows: First, given the path to a Java project, the \texttt{select} module identifies MUTs and corresponding mockable methods within it. The resulting list is configurable by developers.
Next, for each MUT and its mockable methods, the \texttt{instrument} module generates aspects that define the instructions to be followed when their invocations are encountered in production. The aspects are packaged into a \texttt{jar}. This \texttt{jar} need only be attached as a \texttt{javaagent}, while invoking the application as it normally is. The data captured from the invocation of MUTs and mockable methods is serialized and stored to disk. Finally, when triggered by a developer, the \texttt{generate} module processes the data collected for each MUT in production, and generates JUnit tests \cite{junit}, with mocks, stubs, and mock-based oracles defined through Mockito \cite{mockito, 6958396}.

\subsection{Intended use case scenarios}
\rick generates tests based on production observations, and can therefore be used independently of the current state of testing of an application. In fact, the tests generated by \rick can be used to bootstrap a test suite if one does not exist. On the other hand, if the application has a developer-written test suite, the tests generated by \rick can complement it with behaviors that reflect actual usage in production \cite{wang2017behavioral}. Moreover, developers can select MUTs based on custom criteria, such as coverage or mutation testing \cite{9526340}, or target the methods that have recently been modified. The tests generated by \rick mimic production behavior through captured objects and interactions, and can serve as regression tests.

\section{Conclusion}
\rick is an automated test generation tool. Its novelty lies in using data collected from executing applications, to generate tests that mimic production behaviors through mocks, stubs, and mock-based oracles. This demonstration presents an overview of the key phases of \rick, its usage, and use cases. An important future direction is to extend \rick to use finer-grained monitoring and snapshotting techniques, for the generation of tests and mocks. More details on \rick, including evaluations with real-world projects, are available in \cite{tiwari2022mimicking}.

\balance
\bibliographystyle{IEEEtran}
\bibliography{main}

\begin{thebibliography}{10}
\providecommand{\url}[1]{#1}
\csname url@samestyle\endcsname
\providecommand{\newblock}{\relax}
\providecommand{\bibinfo}[2]{#2}
\providecommand{\BIBentrySTDinterwordspacing}{\spaceskip=0pt\relax}
\providecommand{\BIBentryALTinterwordstretchfactor}{4}
\providecommand{\BIBentryALTinterwordspacing}{\spaceskip=\fontdimen2\font plus
\BIBentryALTinterwordstretchfactor\fontdimen3\font minus
  \fontdimen4\font\relax}
\providecommand{\BIBforeignlanguage}[2]{{%
\expandafter\ifx\csname l@#1\endcsname\relax
\typeout{** WARNING: IEEEtran.bst: No hyphenation pattern has been}%
\typeout{** loaded for the language `#1'. Using the pattern for}%
\typeout{** the default language instead.}%
\else
\language=\csname l@#1\endcsname
\fi
#2}}
\providecommand{\BIBdecl}{\relax}
\BIBdecl

\bibitem{mackinnon2000endo}
T.~Mackinnon, S.~Freeman, and P.~Craig, ``Endo-testing: unit testing with mock
  objects,'' \emph{Extreme programming examined}, pp. 287--301, 2000.

\bibitem{thomas2002mock}
D.~Thomas and A.~Hunt, ``Mock objects,'' \emph{IEEE Software}, vol.~19, no.~3,
  pp. 22--24, 2002.

\bibitem{barr2014oracle}
E.~T. Barr, M.~Harman, P.~McMinn, M.~Shahbaz, and S.~Yoo, ``The oracle problem
  in software testing: A survey,'' \emph{IEEE transactions on software
  engineering}, vol.~41, no.~5, pp. 507--525, 2014.

\bibitem{arcuri2015generating}
A.~Arcuri, G.~Fraser, and J.~P. Galeotti, ``Generating tcp/udp network data for
  automated unit test generation,'' in \emph{Proceedings of the 2015 10th Joint
  Meeting on Foundations of Software Engineering}, 2015, pp. 155--165.

\bibitem{islam2010dsc+}
M.~Islam and C.~Csallner, ``Dsc+ mock: A test case+ mock class generator in
  support of coding against interfaces,'' in \emph{Proceedings of the Eighth
  International Workshop on Dynamic Analysis}, 2010, pp. 26--31.

\bibitem{9526340}
D.~Tiwari, L.~Zhang, M.~Monperrus, and B.~Baudry, ``Production monitoring to
  improve test suites,'' \emph{IEEE Transactions on Reliability}, pp. 1--17,
  2021.

\bibitem{demo}
``Rick: Generating mocks from production data,''
  \url{https://youtu.be/ljuUfbC-IZw/}, date accessed 2023-01-06.

\bibitem{graphhopper}
``Graphhopper,'' \url{https://www.graphhopper.com/}, date accessed 2022-12-28.

\bibitem{rick}
``Rick,'' \url{https://github.com/castor-software/pankti#rick}, date accessed
  2022-12-28.

\bibitem{junit}
``Junit 5,'' \url{https://junit.org/junit5/}, date accessed 2022-12-28.

\bibitem{mockito}
``Mockito,'' \url{https://site.mockito.org/}, date accessed 2022-12-28.

\bibitem{6958396}
S.~Mostafa and X.~Wang, ``An empirical study on the usage of mocking frameworks
  in software testing,'' in \emph{2014 14th International Conference on Quality
  Software}, 2014, pp. 127--132.

\bibitem{wang2017behavioral}
Q.~Wang, Y.~Brun, and A.~Orso, ``Behavioral execution comparison: Are tests
  representative of field behavior?'' in \emph{2017 IEEE International
  Conference on Software Testing, Verification and Validation (ICST)}.\hskip
  1em plus 0.5em minus 0.4em\relax IEEE, 2017, pp. 321--332.

\bibitem{tiwari2022mimicking}
D.~Tiwari, M.~Monperrus, and B.~Baudry, ``Mimicking production behavior with
  generated mocks,'' \emph{arXiv preprint arXiv:2208.01321}, 2022.

\end{thebibliography}

\end{document}